\newcommand{\logg}{$\log g$}
\newcommand{\feh}{[Fe/H]}
\newcommand{\teff}{$T_{\rm eff}$}
\newcommand{\kepler}{{\em Kepler}}

\newcommand{\fetwo}{Fe\,{\sc II}}

\documentclass[graybox, footinfo]{svmult}

\usepackage{mathptmx}       
\usepackage{makeidx}         
\usepackage{graphicx}        
\usepackage{multicol}        
\usepackage[bottom]{footmisc}
\usepackage{epstopdf}

\makeindex             

\begin{document}

\title*{'Rapid-fire' spectroscopy of \kepler\ solar-like oscillators}
\author{Thygesen, A. O., Bruntt, H., Chaplin, W. J. \& Basu, S.}
\institute{Thygesen, A. O. \at Zentrum f\"{u}r Astronomie der Universit\"{a}t Heidelberg, Landessternwarte, 69117 Heidelberg, Germany \email{a.thygesen@lsw.uni-heidelberg.de}.
\and
Bruntt, H.\at Department of Physics and Astronomy, Aarhus University, DK-8000 Aarhus C, Denmark \email{bruntt@gmail.com}.
\and
Chaplin, W. J.\at School of Physics and Astronomy, University of Birmingham, Edgbaston, Birmingham B15 2TT, United Kingdom \email{w.j.chaplin@bham.ac.uk}.
\and
Basu, S.\at Department of Astronomy, Yale University, PO Box 208101, New Haven, CT 06520-8101, USA \email{sarbain.basu@yale.edu}.
}
%
%
\maketitle

\abstract{The NASA \kepler\ mission has been continuously monitoring the same field of the sky since the successful launch in March 2009, providing high-quality stellar lightcurves that are excellent data for asteroseismology, far superior to any other observations available at the present. In order to make a meaningful analysis and interpretation of the asteroseismic data, accurate fundamental parameters for the observed stars are needed. The currently available parameters are quite uncertain as illustrated by e.g. \cite{thygesen}, who found deviations as extreme as 2.0 dex in \feh\ and \logg, compared to catalogue values. Thus, additional follow-up observations for these targets are needed in order to put firm limits on the parameter space investigated by the asteroseismic modellers. Here, we propose a metod for deriving accurate metallicities of main sequence and subgiant solar-like oscillators from medium resolution spectra with a moderate S/N. The method takes advantage of the additional constraints on the fundamental parameters, available from asteroseismology and multi-color photometry. The approach enables us to reduce the analysis overhead significantly when doing spectral synthesis, which in turn will increases the efficiency of follow-up observations.}

\section{This Project}
\label{sec:1}
The observations were performed with the FIES instrument on the Nordic Optical Telescope (R=25,000, coverage [3700\AA-7300\AA], S/N$\approx70$). Eight targets were also observed by \cite{bruntt5} (hereafter HB12), using much higher resolution ($R=80,000$) and with a much higher S/N $(\geq200)$. One star (KIC9025370) was found to be a double-lined binary, so this star was discarded in the analysis. The outline of our analysis method follows below.

  \begin{itemize}
	\item Use \logg\ from asteroseismic measurements \cite{basu,gai}.
	\item Calculate \teff\ from TYCHO V and 2MASS K$_S$ photometry using the calibration from \cite{casagrande}, correcting for reddening using the calibration of \cite{munari}.
	\item Calculate microturbulence ($\xi_t$) using calibration from HB12.
	\item Derive initial \feh\ from \fetwo\ lines using the VWA software package \cite{bruntt4}.
	\item Re-iterate \teff\, including metallicity dependence and subsequently redetermine $\xi_t$.
	\item Derive final value of \feh\ from \fetwo\ lines, using final set of fundamental parameters.
  \end{itemize}

\begin{figure}[ht]
\begin{minipage}[b]{0.47\linewidth}
\centering
\includegraphics[width=\textwidth]{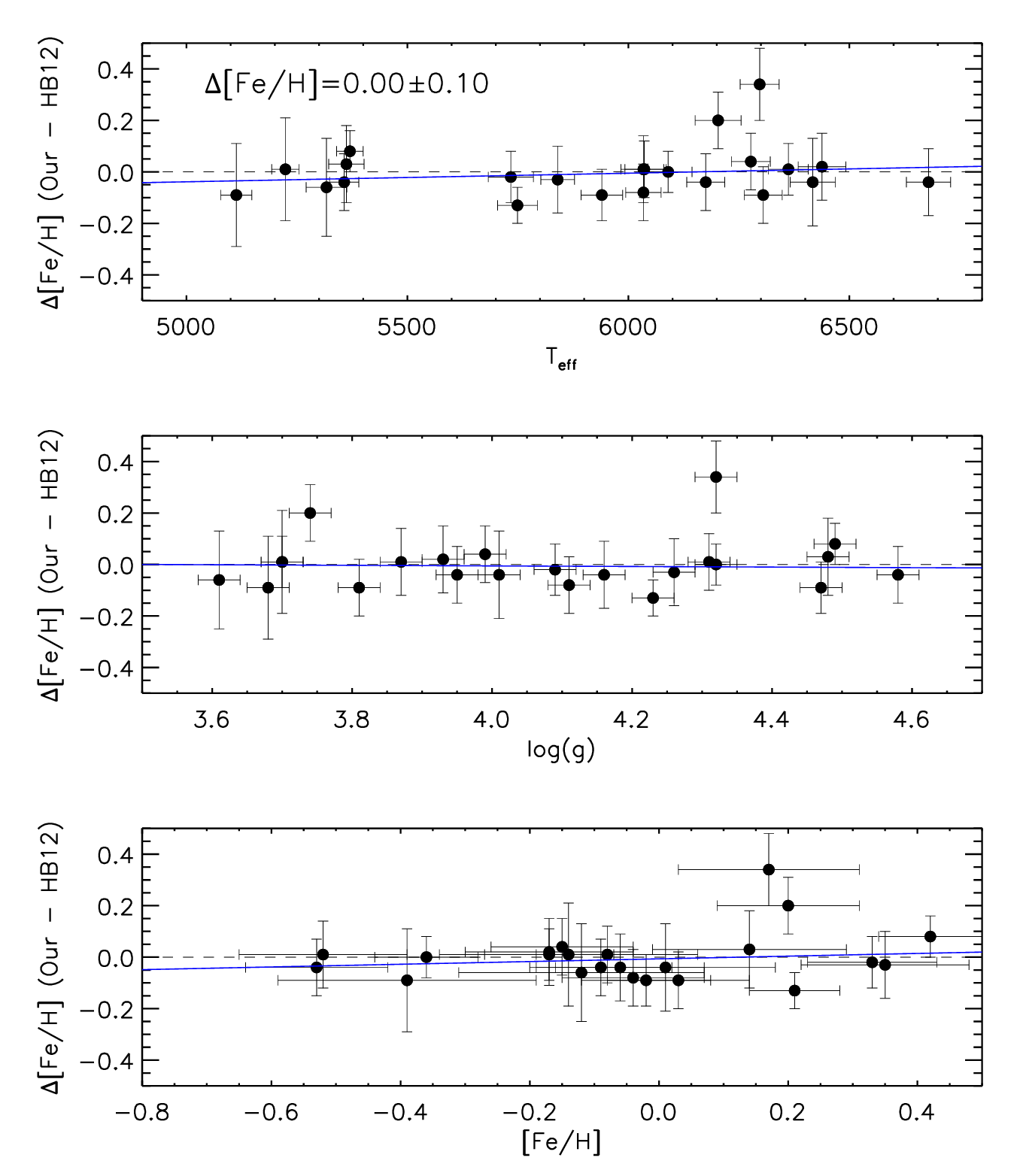}
\caption{Comparison of results from artificially degraded spectra to the results from HB12}
\label{cfht}
\end{minipage}
\hspace{0.5cm}
\begin{minipage}[b]{0.47\linewidth}
\centering
\includegraphics[width=\textwidth]{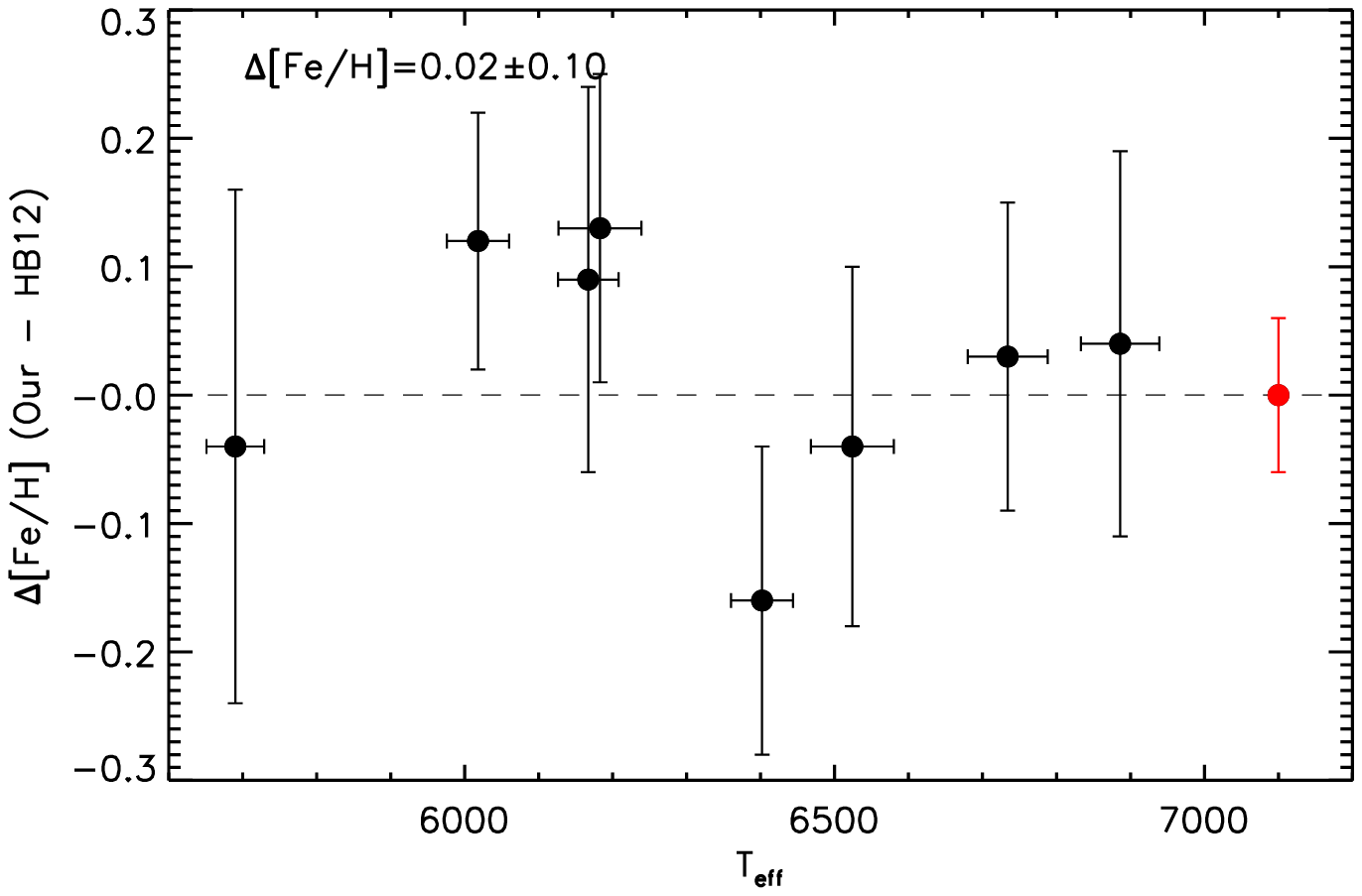}
\caption{Comparison of \feh\ derived from our observations to that of HB12. The red error bar shows the typical uncertainty on \feh\ from HB12.}
\label{reobs}
\end{minipage}
\end{figure}

In Fig.~\ref{cfht} is shown how our analysis of 23 artificially degraded spectra compare to the original results from HB12. The spectra were treated as real observations and analyzed as such. As is evident, we find very good agreement between our analysis and the comparison work, with no offset in \feh\ and an very small scatter. The median uncertainty on our derived metallicities is 0.10 dex, which mainly originates from line-to-line abundance scatter. We observe no clear trends of our metallicity determination with \teff, \logg\ or \feh, suggesting that our method works across the entire parameter space.
      
      In Fig.~\ref{reobs} we present a comparison between the our results for the eight targets which were also observed by HB12 to allow for a direct comparison between different quality observations. Again, the agreement between a dedicated spectroscopic analysis and our proposed method is good, with an insignificant offset and a small scatter. For the observed targets we find a median uncertainty of 0.14 dex on \feh\ which is comparable to the typical uncertainties quoted in standard spectroscopic analyses of higher quality spectra ($\sim0.15$ dex).

With these results, we have illustrated how accurate metallicities can be obtained from moderate quality spectra, when constraints on \teff\ and \logg\ are available from independent sources. This provides a way of quickly deriving metallicities, requiring only measurements of a few \fetwo\ lines and the interstellar Na-D lines, to assess reddening, which will affect the photometric \teff. Before this method can be applied in full, tests of low-metallicity targets as well as fainter targets (V$\geq10$) need to be made, but the results presented here are encouraging. This will help reduce the overheads associated with getting accurate fundamental parameters for all \kepler\ main sequence and subgiant solar-like oscillators.

\begin{acknowledgement}
The authors would like to thank the entire \kepler-team for their continued effort to ensure the success of this mission. AOT acknowledges support from Sonderforschungsbereich SFB 881 "The Milky Way System" (subproject A5) of the German Research Foundation (DFG). SB acknowledges NSF grant AST-1105930. WJC acknowledges the financial support of the UK Science and Technology Facilities Council (STFC). This research took advantage of the SIMBAD and VIZIER databases at the CDS, Strasbourg (France), and NASA's Astrophysics Data System Bibliographic Services. Funding for this Discovery mission is provided by NASA's Science Mission Directorate.
\end{acknowledgement}

\bibliographystyle{spphys}
\bibliography{sesto_proceedings_aot_v2}

\end{document}